\documentclass[acmsmall]{acmart}

\AtBeginDocument{%
  }

\setcopyright{acmlicensed}
\copyrightyear{2026}
\acmYear{2026}
\acmDOI{XXXXXXX.XXXXXXX}

\acmConference[ISSTA '26]{Proceedings of the ACM SIGSOFT International Symposium on Software Testing and Analysis}{October 03--09, 2026}{Oakland, California}
\acmBooktitle{Proceedings of The ACM SIGSOFT International Symposium on Software Testing and Analysis (ISSTA '26), October 03--09, 2026, Oakland, California}

\acmISBN{XXX-X-XXXX-XXXX-X/2026/10}

\usepackage{xspace}
\usepackage{enumitem}
\usepackage{tabularx}
\usepackage{multicol}
\usepackage{multirow}
\usepackage{amsmath}
\usepackage{graphicx}
\usepackage{algorithmic}
\usepackage{algorithm}
\usepackage{booktabs}
\usepackage{makecell}
\usepackage{caption}
\usepackage{colortbl}
\usepackage{tikz}
\definecolor{mytitle}{rgb}{0.9, 0.9, 0.9}
\definecolor{mygray}{rgb}{0.95, 0.95, 0.95}

\renewcommand{\refname}{REFERENCES}

\begin{document}

\title[AMLGuard: Automatic Tracking and Analysis of Crypto Money Laundering via Transaction Semantic Analysis]{Tracing the Shadows: Automatic Tracking and Analysis of Crypto Money Laundering via Transaction Semantic Analysis}

\author{Hao Wu}
\affiliation{%
  \institution{Xi'an Jiaotong University and The Hong Kong Polytechnic University}
  \country{China}
}
\email{emmanuel\_wh@stu.xjtu.edu.cn}

\author{Haijun Wang}
\authornote{Corresponding author}
\affiliation{%
  \institution{Xi'an Jiaotong University}
  \country{China}
}
\email{haijunwang@xjtu.edu.cn}

\author{Shangwang Li}
\affiliation{%
  \institution{Xi'an Jiaotong University}
    \country{China}
}
\email{3124357023@stu.xjtu.edu.cn}

\author{Yin Wu}
\affiliation{
\institution{Xi'an Jiaotong University}
  \country{China}
}
\email{wuyin@stu.xjtu.edu.cn}

\author{Ming Fan}
\affiliation{%
  \institution{Xi'an Jiaotong University}
    \country{China}
}
\email{mingfan@mail.xjtu.edu.cn}

\author{Ting Liu}
\affiliation{%
  \institution{Xi'an Jiaotong University}
    \country{China}
}
\email{tingliu@mail.xjtu.edu.cn}

\author{Xiapu Luo}
\affiliation{%
  \institution{The Hong Kong Polytechnic University}
    \country{China}
}
\email{csxluo@comp.polyu.edu.hk}

\renewcommand{\shortauthors}{H. Wu, H. Wang, S. Li, Y. Wu, M. Fan, T. Liu, and X. Luo}

\begin{abstract}
With the rapid advancement of decentralized finance (DeFi), security incidents related to cryptocurrency have become increasingly prevalent. After such incidents, attackers typically attempt to rapidly move stolen assets, concealing the origin of illicit funds and ultimately converting them into fiat currency.
However, existing anti-money laundering (AML) methods struggle to cope with the semantic complexity of DeFi transactions. They either rely heavily on low-level token transfers, or perform protocol-agnostic money flow analysis, failing to capture the high-level intent of transactions. 
This limitation results in misleading tracing paths with substantial noise and fails when laundering activities span multiple blockchains.

In this paper, we propose AMLGuard, a semantic-aware AML framework for account-based blockchains. 
AMLGuard tracks illicit fund flows from known malicious addresses by performing semantic analysis on complex DeFi transactions, enabling accurate and continuous laundering tracking. Given a complex transaction, AMLGuard combines static rule-based analysis with retrieval-augmented large language model (LLM) reasoning to infer implicit DeFi semantics, transforming raw transaction data into high-level semantic representations.
Furthermore, for cross-chain transactions where laundering intent is not explicitly exposed, AMLGuard parses transaction parameters and performs argument parsing to recover cross-chain semantics, enabling seamless tracking across ledgers.
Based on inferred semantics, AMLGuard abstracts each transaction into a DeFi Semantic Unit (DSU). These DSUs are analyzed and composed iteratively to update account states, expand the tracing frontier, and ultimately construct the illicit fund-flow topology.
 We evaluate the effectiveness of AMLGuard on 82 real-world laundering cases, involving illicit assets worth over $\$$1 billion. Specifically, AMLGuard reconstructs compact illicit fund-flow topologies with destination precision of 94.4\% and 87.6\%, while achieving the highest address recall of 98.4\% and 95.8\% and destination recall of 94.1\% and 93.8\% on single-chain and cross-chain datasets.
 Furthermore, a case study demonstrates that AMLGuard can assist real-world AML investigation, substantially reducing analysis time and effort.
\end{abstract}





\maketitle

\section{Introduction} \label{introduction}

In recent years, blockchain technology has witnessed rapid advancement, laying the foundation for a wide range of decentralized applications. Among these, Decentralized Finance (DeFi) has emerged as a transformative innovation, offering permissionless financial services (e.g., token deposits, lending, and exchanges) without relying on untrusted intermediaries. At the time of writing, 
DeFi protocols managed a total value of $\$$119 billion in digital assets~\cite{DefiLlama}.
However, the inherent openness and pseudonymity of blockchain systems have also made them attractive targets for hackers.
In 2025 alone, blockchain-related security incidents led to losses of about $\$$ 3.35 billion~\cite{web3security}.

In recent years, a wide range of techniques have been proposed to combat illicit activities on blockchain networks.
On the one hand, 
many researchers have applied program analysis techniques to smart contract code~\cite{feist2019slither,wu2024advscanner,sun2024gptscan,wang2020oracle,wu2025rphunter, wang2024skyeye}, enabling the early detection of code vulnerabilities, while others analyze transaction behaviors, aiming to identify suspicious activities in real time~\cite{DeFiRanger,liu2022finding,mazorra2022not,xie2024defort}. 
On the other hand, once an attack has already succeeded and assets have been stolen, the problem fundamentally shifts from detection to incident response, where the core challenge is to trace illicit funds.
In practice, after stealing crypto assets, hackers often engage in money laundering (ML), a process aimed at obscuring the origin of illicit assets. This often involves transferring funds through a series of anonymous addresses, interacting with various DeFi protocols, and eventually cashing out through exchanges. 
At this stage, anti-money laundering (AML) becomes the last and most crucial line of defense against blockchain financial crimes. It plays a crucial role in tracking illicit assets, identifying laundering accounts, and stopping hackers from successfully cashing out.


With the introduction of smart contracts, anti-money laundering on account-based blockchains has become significantly more complex. These blockchain networks enable a wide variety of DeFi protocols, giving rise to highly diverse and complex transaction behaviors. 
In complex DeFi transactions (e.g., aggregator-based token swap), a single external transaction may orchestrate dozens of internal calls across DEX routers and token contracts, resulting in numerous token transfers among multiple addresses.
Existing AML approaches~\cite{wu2023toward,wu2023tracer,lin2024denseflow, huo2025shedding} largely operate at the level of raw token transfer and fail to capture the high-level semantics of such transactions. As a result, they significantly inflate the tracing space, introduce substantial semantic noise, and produce misleading laundering paths.
In practice, public infrastructure contracts (e.g., DEX routers) may be incorrectly identified as laundering participants and recursively traced, diverting attention from truly suspicious entities.
This limitation becomes even pronounced in cross-chain laundering scenarios, where existing solutions ~\cite{yousaf2019tracing,zhang2022cltracer,lin2025connector,lin2025track} frequently fail to establish meaningful semantic associations across blockchain, leading to incomplete broken laundering traces.

Although existing AML methods have demonstrated certain advancements, they still face three challenges. 
\textbf{Challenge 1: Semantic Heterogeneity of DeFi Transactions}. 
In real-world ML scenarios, the behaviors of hacker addresses go far beyond simple token transfers. They frequently interact with a variety of DeFi contracts, each implementing specialized logic for various financial services, inducing highly heterogeneous token movement patterns. Existing methods fail to accurately infer transaction semantics, producing excessive noise and misleading tracing paths.
\textbf{Challenge 2: Implicit Semantics Beyond Deterministic Rules}.
Many critical DeFi behaviors cannot be reliably inferred through deterministic rules alone. Subtle protocol features such as transfer taxes, internal accounting, or multi-stage routing often encode intent implicitly across multiple token movements. Rule-based systems struggle to generalize to such evolving or previously unseen semantics.
\textbf{Challenge 3: Semantic Fragmentation in Cross-chain Laundering}.
The growing adoption of cross-chain protocols introduces a new dimension of semantic discontinuity. Cross-chain operations inherently span multiple ledgers, yet most decentralized bridges do not expose explicit, verifiable links between source-chain and destination-chain transactions. Existing methods fail to recover the underlying cross-chain intent, leading to broken laundering paths.

To address these challenges, we propose AMLGuard, a semantic-aware anti-money laundering framework for account-based blockchains. 
AMLGuard traces suspicious addresses while transforming raw transaction data into high-level semantic representations that guide accurate and continuous laundering tracking.
Given a complex transaction, AMLGuard aims to infer its underlying DeFi semantics, abstracting low-level token transfers into meaningful financial operations that reveal how illicit assets are transformed and where they should be traced next. To address challenge 1,
from an account-centric perspective, AMLGuard introduces a unified semantic abstraction of DeFi operations tailored for AML analysis and applies lightweight static rules to identify explicit DeFi behaviors. To handle implicit and protocol-specific behaviors (challenge 2), it further leverages retrieval-augmented LLM reasoning to infer DeFi operations from structurally similar historical patterns. 
To overcome semantic fragmentation in cross-chain laundering (challenge 3), AMLGuard 
parses transaction parameters and performs argument parsing to identify the corresponding destination-chain address and transaction to be followed, enabling seamless continuation of the laundering trace across ledgers.
Based on the inferred semantics, AMLGuard abstracts each transaction into a DeFi Semantic Unit (DSU), which captures essential DeFi behaviors while filtering out irrelevant execution noise. These DSUs are analyzed and composed iteratively to expand the tracing frontier and construct the illicit money-flow topology.

To evaluate the effectiveness of AMLGuard, we curated a dataset of
82 real-world money laundering incidents involving over \$ 1 billion in illicit assets, including 63 single-chain ($D_s$) and 19 cross-chain ($D_c$) cases. 
On average, AMLGuard traces 299 transactions and 44 labeled addresses per incident on $D_s$, and 287 transactions and 74 addresses on $D_c$, while achieving high destination precision (94.4\% and 87.6\%).
Despite maintaining compact illicit fund-flow topologies with limited false positives, AMLGuard achieves the highest address recall of 98.4\% and 95.8\% and destination address recall of 94.1\% and 93.8\% on two datasets, respectively, outperforming existing methods.
Finally, we evaluated the practical benefit of AMLGuard in real-world AML investigation through case studies, showing that it can reduce auditing time and improve the accuracy of laundering tracing. A prototype of AMLGuard, our datasets and experimental results are available online~\cite{AMLGuard}.


In summary, the contributions of this paper are as follows:

\begin{itemize}[leftmargin=*] 
    \item We propose a semantic-aware anti-money laundering framework AMLGuard capable of accurately analyzing transaction semantics, tracing cross-chain money laundering activities, and producing compact illicit money-flow topology.
    \item To handle complex DeFi transactions, we design a semantic analysis module combining static rules with LLM reasoning to abstract low-level transaction data into high-level DeFi Semantic Units. Furthermore, to support cross-chain laundering, we introduce a cross-chain semantic recovery mechanism that infers cross-chain intent, enabling seamless money laundering trace.
    \item We have conducted a comprehensive evaluation of AMLGuard using 82 real-world AML incidents. The results show that AMLGuard reconstructs compact illicit fund-flow topologies with destination precision of 94.4\% and 87.6\%, while achieving the highest address recall of 98.4\% and 95.8\% and destination recall of 94.1\% and 93.8\% on single-chain and cross-chain datasets.
\end{itemize}

\section{Background}

In this paper, we focus on blockchain networks running on the Ethereum Virtual Machine (EVM), which adopt an account-based model. In this model, an account typically refers to an entity that holds digital assets like cryptocurrencies or tokens on blockchain networks. Accounts can be categorized into two types: Externally Owned Accounts (EOAs), which are controlled by private keys
, and Contract Accounts (CAs), known as smart contracts, which are self-executing programs with predefined logic.
There are two types of transaction: external transactions, which are initiated by EOAs, and internal transactions, which are triggered during the execution of smart contracts. 


\subsection{Decentralized Finance}
Decentralized Finance (DeFi) offers a wide range of financial services without relying on centralized intermediaries. DeFi protocols are typically implemented as smart contracts deployed on blockchain platforms. These services are powered by crypto assets, which serve as the medium for value transfer. They can generally be categorized into two types: native tokens (e.g., ETH on Ethereum) and tokens, which are custom assets created and managed through smart contracts. These tokens follow specific token standards that define their functionality and interoperability, e.g., ERC-20 and ERC-721. 
Decentralized Exchange (DEX) is a peer-to-peer marketplace, which allows users to swap tokens directly from their wallets without intermediaries. For instance, hackers can use Uniswap, one of the most widely used DEXs, to convert illicit tokens (e.g., DAI) into more liquid assets like ETH, facilitating the next stages of money laundering.
 Beyond exchanges, DeFi protocols also offer lending and staking services. Users can deposit assets as collateral to borrow other tokens or stake funds to earn rewards, mimicking traditional financial mechanisms in a decentralized manner.

\subsection{Cross-chain Bridge} \label{cross-chain}
Cross-chain bridges can be broadly categorized into centralized (CeFi) and decentralized (DeFi) designs. CeFi bridges rely on custodial EOAs and internal ledgers for accounting, whereas DeFi bridges implement the bridging logic entirely through smart contracts. In this work, we focus on the latter because decentralized bridges typically do not provide a verifiable linkage between source and destination chain transactions, making them a major challenge for AML tracing.

A typical DeFi cross-chain bridge consists of three components: the source chain part, the cross-chain layer, and the destination chain part. The source and destination chain parts host smart contracts that manage asset locking and minting, while the cross-chain layer is composed of off-chain relayers responsible for information propagation. With the cross-chain bridge, users can deposit assets on the source chain and withdraw corresponding assets on the destination chain. Here, we use a simplified example to illustrate the workflow of a cross-chain asset transfer: 
\begin{enumerate}[leftmargin=*] 
    \item The user invokes the cross-chain bridge contract on the source chain and sends the corresponding assets. The bridge contract locks the assets and emits an event.
    \item  Off-chain relayers monitor these events, verify the lock operation, and relay the validated message to the destination chain bridge contract.
    \item The cross-chain bridge contract on the destination chain verifies the relayed information and releases (or mints) the bridged assets to the user-specified address.
\end{enumerate}

\subsection{Anti-money Laundering}

The goal of anti-money laundering on blockchain is to track the flow of illicit funds from a designated hack address, construct the illicit fund-flow topology, and determine the final destination of tainted funds.
Similar to the three-phase model of money-laundering in traditional finance~\cite{cassella2018toward}, we define the three stages of the crypto ML process:

\begin{enumerate}[leftmargin=*] 
    \item \textbf{Placement Phase:} The attackers execute an exploit and place the stolen funds to attacker-controlled addresses. 
    \item \textbf{Layering Phase:} During this phase, the illicit assets are moved through a complex series of transfers. Attackers perform multi-hop transfers, interactions with DeFi protocols (e.g., swap, lending, and staking), or movements across cross-chain bridges to obscure the origin of funds and increase laundering complexity.
    \item \textbf{Integration Phase:} Eventually, the laundered assets are
 aggregated into exit services such as CEXs or privacy-enhancing mixers for final cash-out or further obfuscation.
\end{enumerate}

In this work, we focus primarily on tracking laundering flows up to these exit services, most notably CEXs, where regulatory collaboration can help freeze illicit funds and trigger off-chain investigations via Know-Your-Customer (KYC) procedures. And tracing funds after entering privacy-enhancing mixers is beyond the scope of this study, we provide further discussion in Section~\ref{discussion}.

\section{Motivating Example}

\begin{figure}[htbp]
\centerline{\includegraphics[width=0.66\linewidth]{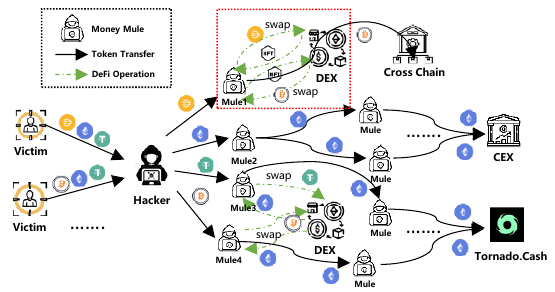}}
\caption{Money Laundering Process of Cream Finance Incident}
\label{Money Laundering Process of Cream Finance}
\vspace{-0.3cm}
\end{figure}

On Oct. 28th, 2021, \textit{Cream Finance}, a DeFi lending platform on Ethereum, suffered a devastating price manipulation attack, resulting in \$130 million losses~\cite{CreamFinance}. 
A simplified illustration of the hacker's subsequent money laundering process is shown in Fig.~\ref{Money Laundering Process of Cream Finance}. 
After the exploit, the hacker extracted a portfolio of stolen assets (e.g., ETH, DAI, LRC) and initiated the laundering by distributing the stolen funds to multiple mule addresses. Given the limited fiat liquidity of certain tokens on blockchain networks, these mule addresses utilized various DEXs to convert the stolen tokens into more liquid cryptocurrencies, primarily ETH. 
The laundering continued through a multi-layer obfuscation process involving 31 addresses across four layers of transfers, culminating in the integration of funds into coin mixers, cross-chain bridges and CEXs.


To concretely illustrate the challenges of on-chain AML, we examine the behaviors of mule addresses and take Money Mule 1 as an example~\cite{hackaddress}, shown in Fig.~\ref{Money Laundering Process of Cream Finance}.
Upon receiving a large amount of DAI token from the hacker’s primary address, Money Mule 1 invoked the \textit{$swap\_into\_synth$} function of the SynthSwap contract~\cite{stake-transaction}.
This operation involved staking approximately 2 million DAI to mint a synthetic asset, represented as a CRV/SS ERC-721 NFT, which served as a proof-of-stake certificate. 
Next, Money Mule 1 redeemed the synthetic NFT, converting it into renBTC~\cite{redeem-transaction} and then interacted with a cross-chain contract, executing a cross-chain transfer that further obfuscated the origin of the stolen assets.

Taking the stake transaction as an example, the single external transaction triggered 134 internal transactions and resulted in 9 token transfers across 7 distinct addresses, resulting in highly complex behaviors. 
However, from a token-centric perspective, its underlying intent is simply a token exchange. Existing AML methods fail to recognize this high-level semantics and instead treat intermediate token movements as laundering paths, introducing substantial noise and redundant branches. Protocol-agnostic money-flow analysis further risks misclassifying protocol-internal tax or accounting addresses as laundering participants, leading to incorrect traces. This issue is exacerbated when the assets enter cross-chain protocols: the source-chain transaction records only a bridge contract invocation, without any explicit on-chain linkage to the destination-chain transaction, making it difficult for existing methods to continue tracking the laundering flow. 

This example underscores the importance of transaction semantic analysis in on-chain AML. By abstracting low-level token transfers into high-level DeFi operations, semantic analysis filters protocol-internal noise, correctly captures how illicit assets are transformed, and guides accurate continuation of tracing.

\section{Methodology}

\subsection{Overview of AMLGuard}

\begin{figure*}[htbp]
    \centerline{\includegraphics[width=\linewidth]{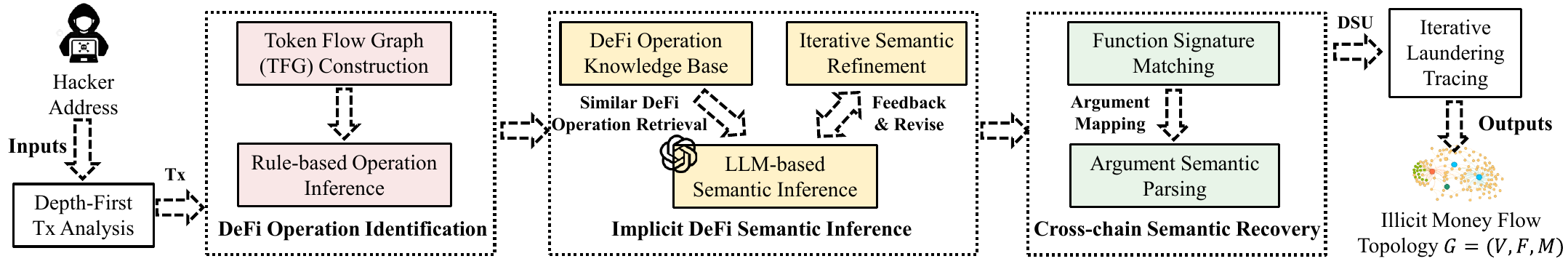}}
    \caption{Overview of AMLGuard}
    \label{Overview of AMLGuard}
    \vspace{-0.2cm}
\end{figure*}

We introduce AMLGuard, an iterative framework to track crypto money laundering through transaction semantic analysis. 
As shown in Fig.~\ref{Overview of AMLGuard}, AMLGuard progressively reconstructs the illicit money-flow topology by analyzing each transaction and using its inferred semantics to guide subsequent exploration.
Starting from a known hacker address, AMLGuard retrieves its transaction histories and performs the depth-first tracing process. 


Specifically, for each transaction, AMLGuard constructs a Token Flow Graph (TFG) to capture token movements. Based on the TFG, from an account-centric perspective, AMLGuard defines a set of well-known DeFi operations and applies static graph search rules to infer explicit DeFi operations. To handle implicit and protocol-specific operations, AMLGuard retrieves similar TFG structures and their associated semantics from a curated knowledge base, and employs LLM-based reasoning in a retrieval-augmented (RAG) pipeline to infer the most plausible DeFi operation. The inferred intent is further validated and, if necessary, refined to ensure semantic correctness.
For transactions involving cross-chain interactions, AMLGuard further analyzes transaction parameter and parses argument semantics to identify the corresponding destination-chain address and transaction to be traced. Based on the inferred semantics, AMLGuard abstracts each transaction into a DeFi Semantic Unit (DSU), which captures the essential financial behavior while filtering out low-level execution noise. These DSUs are iteratively composed to update account states, expand the tracing frontier, and ultimately construct the illicit money-flow topology.

\subsection{DeFi Operation Identification}\label{semantic analysis}
Raw blockchain transactions expose execution details but obscure the underlying financial intent, creating a semantic gap between low-level traces and high-level DeFi behaviors. 
To bridge this gap, AMLGuard performs semantic lifting by modeling token movements as a directed Token Flow Graph (TFG), from which transactions are abstracted into high-level financial behaviors.
According to the study~\cite{zhong2025defiscope}, only a few DeFi operations span multiple transactions. Therefore, this paper focuses solely on the semantic analysis of single-transaction granularity.

\subsubsection{Token Flow Graph Construction}

Given a target address, we retrieve all its external transactions from blockchain. For each transaction (termed as tx), we further collect: (1) internal native-token transfers triggered during execution, and (2) token transfer events emitted in execution logs, including both ERC20 and ERC721 tokens.
For these three types of token transfer actions (i.e., raw tx, internal txs, and token transfer events), 
we define each such movements as a \textit{transfer action}, capture the sender, receiver, token, amount, and execution order within the transaction.

\begin{definition}[\textsc{Transfer Action}]
    \textit{A transfer action $Ta<$s, r, t, a, i$>$ denotes that account s transfers the amount a of the token t to account r at time index i.} 
\end{definition}

Based on these \textit{transfer actions}, we construct a \textit{Token Flow Graph} to model intra-transaction token movements.
In the TFG, nodes represent participant accounts, including externally owned accounts (\textit{EOAs}) and contract accounts (\textit{CAs}), collectively referred to as participant accounts (\textit{PAs}). Directed edges encode token transfers between PAs. Multiple edges may exist between the same node pair, differentiated by token type, transfer amount, and execution index.

\begin{figure*}[htbp]
        \centering
        \includegraphics[width=0.75\textwidth]{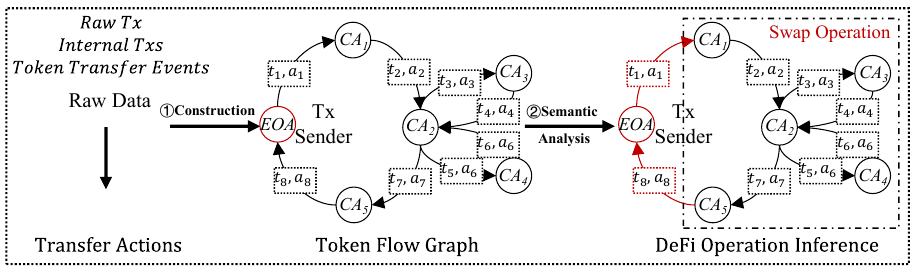}
        \caption{An Example Showing the Workflow of Transaction Semantic Analysis}
        \label{fig:left}
        \vspace{-0.4cm}
\end{figure*}


\begin{definition}[\textsc{Token Flow Graph}]
    \textit{A token flow graph TFG $::=<$$\mathbb{V}$, $\mathbb{E}$$>$, where $\mathbb{V}$ is the set of accounts (including EOAs and CAs), $\mathbb{E}$ is the set of directed edges, i.e., $\mathbb{E} = \{E_{1}, ..., E_{m}\}$, where each $E_{n} ::= <T_{n}.t, T_{n}.a, T_{n}.i>$, $T_{n} \in \mathbb{T}$, $\mathbb{T}$ is the set of transfer actions involved in the raw tx.}
\end{definition}

Fig.~\ref{fig:left} illustrates a token swap operation from our motivating example.
In this example, the tx sender first transfers the tokens to the \textit{$CA_{1}$}, which subsequently interacts with multiple intermediate contracts. Eventually, the swapped tokens are transferred back to the sender.
Such transactions involve multiple token transfers and complex contract interactions, making it difficult to directly trace illicit fund conversions and movements from raw transaction traces.

\subsubsection{Rule-based DeFi Operation Identification}

To systematically model DeFi operations relevant to anti-money laundering analysis, we conduct an in-depth study of DeFi application types listed on DeFiLlama~\cite{DefiLlama}, a widely used DeFi analytics platform. Specifically, we focus the top 15 DeFi application types that collectively account for over 95\% of the ecosystem's Total Value Locked (TVL). 
Our analysis reveals that, despite their seemingly diverse functionalities and business logics, 
most DeFi applications exhibit highly similar financial behaviors from a token-flow perspective. Conceptually, they resemble traditional banking systems, aggregating liquidity through deposits, redistributing liquidity via swaps or lending, extracting values through fees or liquidation, and returning assets or profits to users.
From an token-centric viewpoint, the seemingly diverse financial behaviors can be distilled into a small set of fundamental interactions with DeFi applications. 

Motivated by this observation, we introduce a unified conceptual abstraction of DeFi operations tailed for the AML analysis. 
Rather than explicitly modeling user intent, our abstraction characterizes transaction by how tokens are transformed, transferred or exchanged across accounts and contracts. Based on this principle, we define five core DeFi operations that form the foundation for transaction semantic analysis in our system. (1) \textbf{Approve}: the authorization process for token transfers, whereby an account grants another account or contract permission to spend a specified amount of tokens; (2) \textbf{Transfer}: the transfer of native tokens or ERC tokens between accounts, including NFT mint/burn and token distributions such as airdrops; (3) \textbf{Swap}: the exchange of one type of token for another type of token or proof-of-ownership token. 
Beyond DEXs, it also subsumes liquidiy-related behaviors such as adding or removing liquidity,as well as certain deposit or withdrawal interactions that result in token-type transformation.
(4) \textbf{Deposit}: the action of depositing tokens into smart contracts,commonly observed in staking or yield-farming protocols; (5) \textbf{Withdraw}: the retrieval of previously deposited tokens, including reclaiming staked assets or receiving borrowed tokens in lending protocols. 

\begin{table*}[htbp]
\centering
\vspace{-0.15cm}
\caption{DeFi Operation Identification Rules and Semantic Units}
\vspace{-0.1cm}
\small
\scalebox{0.78}{
\begin{tabular}{>{\arraybackslash}p{2.15cm}  >{\arraybackslash}p{8.8cm}  >{\arraybackslash}p{5.7cm}} 
\toprule

 \textbf{DeFi Operation} & \textbf{Identification Rules} & \textbf{Semantic Units}  \\
\midrule
  Approve &  $Ap(OA_{owner},PA_{spender},t_1,a_1,i_1)$   & $O_{Approve}<OA,PA,(t_1,a_1,\__),(t_1,a_1,\__)>$ \\
\midrule
 Transfer & $Ta(OA_1,OA_2,t_1,a_1,i_1)$  &    $O_{Transfer}<OA_1,OA_2,(t_1,a_1,\__),(t_1,a_1,\__)>$ \\
\midrule
 Swap & $Ta(OA,A_i,t_1,a_1,i_1) \wedge Ta(A_i,A_j,t_2,a_2,i_2) \wedge...\wedge Ta(A_k,PA,t_n,a_n,i_n)$  &  $O_{swap}<OA,PA,(t_1,a_1,\__),(t_n,a_n,\__)>$\\ 
 \midrule
 Deposit & $Ta(OA,A_i,t_1,a_1,i_1) \wedge Ta(A_i,A_j,t_1,a_1,i_2) \wedge...\wedge Ta(A_k,CA,t_1,a_1,i_n)$   & $O_{Deposit}<OA,CA,(t_1,a_1,\__),(t_1,a_1,\__)>$ \\
 \midrule
 Withdraw & $Ta(A_i,A_j,t_1,a_1,i_1) \wedge Ta(A_j,A_k,t_1,a_1,i_2) \wedge...\wedge Ta(A_k,OA,t_1,a_1,i_n)$   & $O_{Withdraw}<A_i,OA,(t_1,a_1,\__),(t_1,a_1,\__)>$ \\
\bottomrule
\end{tabular}}
\label{table:inference rules}
\vspace{-0.1cm}
\end{table*}



To infer DeFi operation from the TFG, we design a set of operation-specific graph search procedures over on directed TFG. 
The algorithm initiates graph traversal from the tx sender, explores token flow paths and matches candidate subgraphs against predefined identification rules, shown in Table~\ref{table:inference rules}.
We use the \textit{swap} operation as a representative example. Specifically, we perform a depth-first search (DFS) on the TFG to detect cyclic token-flow paths that originate and end at EOAs, where the token types of the first and the last transfer edges differ. And the time index of the traversed edges must be monotonically increasing.
If such a subgraph is detected, it is identified as a swap operation. In practice, the account providing the input tokens and the account receiving the output tokens may differ.
To ensure the semantic integrity of the match, the terminal account in the path is either a leaf node or a node whose outgoing edges have already been fully explored during the search. This constraint prevents premature or ambiguous matches and ensures that the inferred subgraph captures a complete semantic unit.

As shown in Fig.~\ref{fig:left}, once a swap operation is inferred, the intermediate token transfers within the rectangular area can be treated as low-level execution details. 
The essential semantic signal lies in the tokens sent and received by EOAs, which jointly characterize the swap behavior. Accordingly, we abstracts the set of fine-grained \textit{transfer action} into a single DeFi operation, reducing transaction complexity and providing an interpretable representation for downstream AML tracing.

\subsection{Implicit DeFi Semantic Inference}\label{llm_based}

While the rule-based method reliably identifies many standard DeFi operations, a subset of transactions remains challenging to classify. 
In practice, protocol-internal accounting behaviors--such as fee redistribution or tightly coupled composite operations--often introduce token movements that deviate from canonical patterns.
To address such ambiguous cases, we leverage the in-context reasoning ability of LLMs to infer the most plausible DeFi semantics from noisy token-flow structures, enabling robust semantic analysis beyond rigid graph search rules.

\subsubsection{Retrieval-Augmented Operation Infernce}

\begin{figure*}[htbp]
        \centering
        \includegraphics[width=0.94\textwidth]{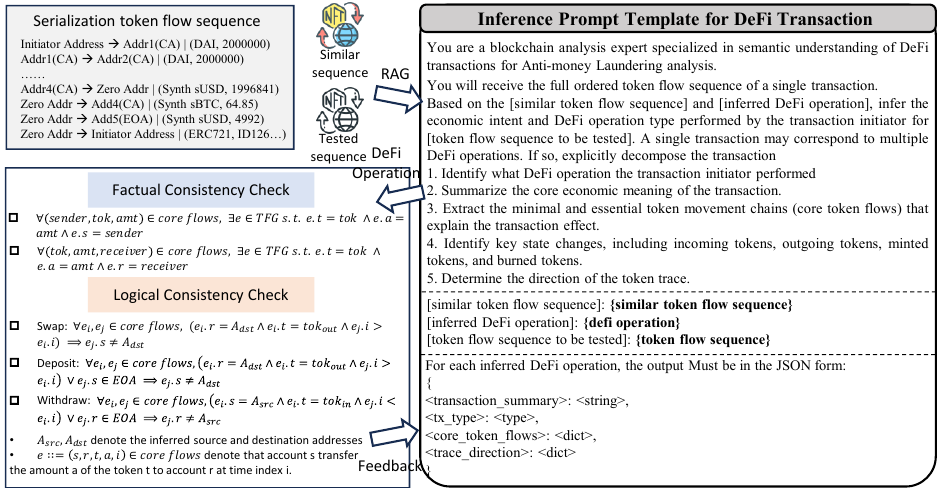}
        \caption{LLM-based DeFi Semantic Inference}
        \label{llm}
        \vspace{-0.2cm}
\end{figure*}

We adopt a RAG pipeline to ground LLM-based semantic inference in domain knowledge. The core idea is to expose the LLM to previously analyzed DeFi operations with similar token-flow structures, enabling analogy-driven reasoning. To this end, we construct a DeFi Operation Knowledge Base by serializing TFGs and computing their embeddings for similarity search. During serialization, we normalize address representations by labeling contract types, removing irrelevant identifiers, and replacing token addresses with standardized token symbols, as shown in Fig.~\ref{llm}. 
For each entry, we curate a concise semantic annotation that summarizes key token state transitions, the DeFi operation, and its implications for subsequent tracing. This results in a structured mapping between token-flow patterns and high-level semantics.
The details are shown in Section~\ref{experiment}.

Given a complex TFG to analyze, AMLGuard applies the same serialization and embedding procedure to retrieve the most similar reference operations and their semantic annotations from the knowledge base. 
These retrieved examples are then incorporated as few-shot demonstrations in a structured prompt that guides the LLM in inferring the DeFi semantics. 
As illustrated in Fig.~\ref{llm}, the prompt template consists of: (1) inference instructions defining the expected reasoning behavior, (2) the retrieved reference operations alongside the target TFG, and (3) a constrained output schema. 
Conditioned on these contextual examples, the LLM infers the most plausible DeFi operation or operation composition corresponding to the observed token flow pattern. 



\subsubsection{Iterative Semantic Validation and Refinement}
The inference process is performed iteratively to ensure both robustness and semantic correctness. First, outputs that do not conform to the predefined schema are automatically rejected and regenerated. More importantly, to mitigate hallucinations and enforce semantic reliability, we incorporate an operation correctness check that validates the inferred DeFi semantics against the observed token movements in the TFG.

The module enforces two complementary forms of consistency.
\textbf{Factual consistency} verifies that the inferred semantic description is grounded in the TFG, ensuring that all claimed core token flows correspond to actual graph edges, and that referenced token symbols and transferred amount are compatible with recorded transfer actions.
\textbf{Logical consistency} evaluates whether the inferred operation satisfies a set of operation-specific feasibility constraints, which encode necessary structural and semantic properties of valid DeFi Operation (Fig.~\ref{llm}). Importantly, these constraints are not intended to uniquely identify or confirm a specific DeFi operation. Instead, they are designed to invalidate implausible or inconsistent inferences produced by the LLM. 
For example, in a Swap operation, let the inferred core token flow have $out\_token:tok_1$, $out\_address:a_1$, and occur at $time\_index:i_n$. One swap feasibility constraint requires that once address $a_1$ receives Token $t_1$ as core flow of the swap operation, it must not subsequently act as a sender of token $tok_1$ at any time index $i>n$. The presence of such post-swap outgoing transfers of token $tok_1$ from address $a_1$ violates swap semantics and is flagged as a logical inconsistency, which triggers revision and enables the system to correct erroneous inferences.

When a check fails, the module produces concise and structured diagnostics that are fed back into a revision prompt to iteratively refine the inference. 
The revision prompt follows the same structure as the inference prompt and is therefore omitted for brevity. 
In this stage, we explicitly instruct the LLM to (1) interpret the reported inconsistencies, and (2) revise its inference accordingly, reusing the same retrieved reference example. To prevent infinite correction loops, a maximum iteration limit is imposed. 
Overall, the iterative inference-and-verification workflow substantially reduces hallucinations, improves semantic fidelity, and yields more reliable transaction semantics for downstream tracing and state-updating.


\subsection{Cross-Chain Semantic Recovery}
Although above transaction semantic analysis enables the interpretation of intra-chain behaviors, cross-chain protocols inherently fragment money-flow continuity, introducing blind spots that impede end-to-end laundering tracing. 
To overcome this limitation, we propose a cross-chain semantic recovery mechanism that reconstructs cross-chain intent from source-chain evidence. Our approach combines two complementary techniques, i.e., 1) matching function signature templates and 2) utilizing argument semantic analysis.

\begin{figure*}[htbp]
        \centering
        \includegraphics[width=0.65\textwidth]{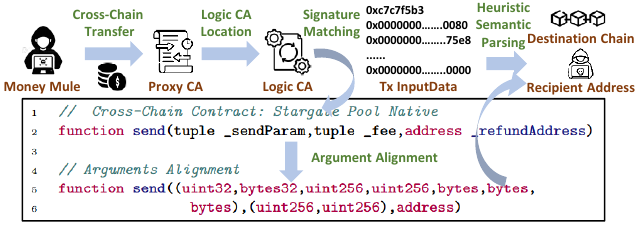}
        \caption{An Example Showing the Workflow of Cross-Chain Semantic Recovery}
        \label{crosschain}
        \vspace{-0.3cm}
\end{figure*}

In laundering scenarios, attackers typically initiate cross-chain requests on the source chain by transferring assets into protocol-controlled vault contracts. Semantically, such actions correspond to \textit{Deposit} DeFi operations. We systematically inspect all inferred \textit{Deposit} transaction semantics and match the invoked contract functions against a curated set of cross-chain function signature templates collected from Chainspot~\cite{chainspot}, based on their official documentation. Once a signature match is identified, the associated function arguments are decoded to recover essential cross-chain semantics, such as the destination chain and the intended recipient address. In the remainder of this section, we focus on the second technique--argument semantic analysis--and detail how AMLGuard leverages protocol-specific calldata semantics to recover cross-chain intent and enable continuous laundering tracking across heterogeneous blockchains. An example is shown in Fig.~\ref{crosschain}.

\subsubsection{Function Signature Matching}
In cross-chain transactions, the sender typically encodes transaction metadata, such as the destination chain and the recipient address, into the \textit{inputdata} field. However, \textit{inputdata} is a raw hexadecimal byte sequence, where the first four bytes specify the function selector and the remaining bytes encode arguments, rendering cross-chain semantics opaque at the execution level.
To recover these semantics, we first decode the \textit{inputdata} by resolving the invoked function and its parameter layout. 
Specifically, we leverage the Web3 library to retrieve the ABIs of the target contract and reconstruct all candidate function signatures. By matching the extracted function selector against these signatures, we precisely determine the invoked function and its associated parameters, thereby exposing the semantic structure of the cross-chain request.

In practice, many cross-chain protocols adopt proxy-based architectures, where users interact with proxy contracts rather than the underlying logic contracts that implement cross-chain functionality. As proxy contracts often do not expose the ABI of the delegated logic, direct signature resolution is often feasible. To address this, we implement a proxy resolution mechanism that identifies the underlying logic contracts by inspecting contract storage layouts.
AMLGuard supports three widely adopted proxy patterns: EIP-1967~\cite{EIP1967}, EIP-1967 beacon~\cite{EIP1967}, and EIP-2535 diamond~\cite{EIP2535}.
Once the logic contract is resolved, we retrieve its ABI sets and apply the same signature-matching procedure. This mechanism enables reliable identification of cross-chain function invocations and provides a necessary foundation for subsequent argument semantic analysis.

\subsubsection{Argument Semantic Parsing}

After resolving the invoked cross-chain function and its ABI, we perform argument semantic parsing to recover cross-chain intent from decoded calldata. Specifically, we align each function argument with its runtime value in the \textit{inputdata} field according to the ABI-defined type system, producing a strcutured mapping between argument names, types and values.
We categorize function arguments into three classes and process them accordingly:
\begin{enumerate}[leftmargin=*]
    \item Primitive type (e.g., address and uint) are directly mapped to their decoded values.
    \item Opaque types include string and bytes. String arguments are analyzed to identify embedded addresses or protocol-specific annotations, while bytes arguments are examined as raw byte sequences to extract candidate EVM addresses or other encoded identifiers.
    \item Structured types (e.g., tuples and arrays) are recursively flattened, ensuring that all nested fields are normlized into name-type-value triples.
\end{enumerate}

Based on the normalized argument mapping, we perform heuristic semantic parsing to recover key cross-chain intents.
Destination chains are inferred when integer values corresponding to known chain identifiers and their argument names contain chain-related keywords (e.g., chain, id). 
Recipient addresses are identified from address-type fields, or extracted from string and bytes arguments containing valid address patterns whose names also match predefined recipient-related terms. 
To reduce noise, parameters associated with non-routing roles (e.g., tokens, senders) are excluded. 
Through semantic parsing, we obtain the destination chain identifier and recipient address encoded in the transaction. 
These semantics are then used to refine the corresponding \textbf{Deposit} operation and transform it into a \textbf{Transfer} operation, enabling end-to-end tracking.

\subsection{Semantic-guided Laundering Tracing}\label{account state}


\subsubsection{DeFi Semantic Unit Abstraction}
Building on the semantic analysis described above, our goal is to distill each transaction into a representation that captures (1) the initiator's actionable intent, (2) the resulting token-state transitions, and (3) the next-hop direction for laundering-trace expansion. To achieve this, we introduce the \textit{DeFi Semantic Unit} (DSU), a unified abstraction that standardizes heterogeneous DeFi transactions, serving as the atomic semantic element for laundering tracing.

\begin{definition}[\textsc{DeFi Semantic Unit}]
    \textit{A DeFi Semantic Unit $O_{T}$ ::= $<\mathbb{S},\mathbb{R},\mathbb{S}_t,\mathbb{R}_t>$, where T represents the type of DeFi operations, $\mathbb{S}$ is the logical initiator of the operation, $\mathbb{R}$ is the target entity of the operation, and $\mathbb{S}_t/\mathbb{R}_t ::=<t_s,a_s,c_s> or <t_r,a_r,c_r>$ denotes the token type, amount, and corresponding blockchain associated with the initiator or target (e.g., token approved or transferred).}
\end{definition}

Each DSU encapsulates the intent and effect of a single operation, while abstracting away low-level execution details. First, it enables quantitative analysis of each account involved in the transaction. Second, it determines which entities warrant further tracing and which can be safely pruned. 
For example, if an account $A$ exchanges one token for another and the output token is received by account $B$. The resulting swap DSU focuses solely on the token balance changes of $A$ and $B$. The DEX contract and its auxiliary addresses are excluded from further analysis. 
For cross-chain transaction, the initial DSU is typically inferred as a \textbf{Deposit} operation on the source chain. After cross-chain semantic recovery, AMLGuard refines this DSU by updating the receiver address and chain attribute, transforming it into a \textbf{Transfer} DSU. This refinement restores semantic continuity across chains and enables end-to-end laundering tracing.
This abstraction prunes irrelevant paths and substantially reduces tracking overhead while preserving all laundering-relevant information.

Then, AMLGuard parses the token semantics encoded in each DSU and updates the states of all involved accounts accordingly. 
Given a \textit{DSU} that encodes token transfer, 
AMLGuard extracts the sender \textit{s} and receiver \textit{r} along with associated token flow information $<t_s, a_s, c_s>$ and $<t_r, a_r, c_r>$. The \textit{Account State} for account \textit{s} and \textit{r} are then deterministically updated. Specifically, for the sender and receiver, AMLGuard deducts and increases the token amount for the token balance mappings respectively. 
Meanwhile, the current \textit{DSU} is appended to the operation history of account, preserving a chronological record of semantically meaningful interactions for each account.

\subsubsection{AML Workflow}

AMLGuard formulates anti-money laundering as an iterative, semantic-guided tracing process. Algorithm~\ref{aml} summarizes the workflow which takes an initial hacker address and a specified block number as input, and outputs the laundering topology together with the final destinations of illicit asset. 

The algorithm initializes a stack-based address set $ST_{a}$ with the seed hacker address and an empty visited transaction map $VisitedTxMap$ to record analyzed transactions for each address (L1-L2). 
It then performs depth-first traversal over addresses in $ST_{a}$. 
At each iteration, the address at the top of the stack is selected as the current analysis target \textit{ca} (L3) and its previously unanalyzed transactions are processed sequentially (L4). 
Each transaction $t$ is semantically analyzed to infer its corresponding \textit{DeFi Semantic Unit} (DSU) via \textit{SemanticAnalyze(t)} (L5). If the transaction involves a cross-chain interaction, cross-chain semantic recovery is invoked to reconstruct the transaction intent and the recovered destination-chain and recipient information is incorporated into DSU (L6).
The inferred DSU is then used to deterministically update the states of involved accounts (L7) and the transaction is marked as visited to prevent redundant analysis in subsequent iterations (L8).


\begin{algorithm}[htbp]
\caption{Semantic-Driven Money Laundering Tracking}
\label{aml}
\footnotesize
\begin{algorithmic}[1]
\REQUIRE Seed address $h$, start block $bn$
\ENSURE Illicit money-flow topology $G$, Laundering destinations $Des$
\STATE Initialize address stack $ST_a \gets \{h\}$ and visited map $VisitedTxMap \gets \emptyset$
\WHILE{$ST_a \neq \emptyset$}
    \STATE $ca \gets ST_a.peek()$
    \FOR{each unanalyzed transaction $t$ of $ca$}
        \STATE $DSU \gets SemanticAnalyze(t)$
            \hfill \texttt{ // Perform Transaction Semantic Analysis}
        \STATE $DSU \gets CrossChainRecover(t, DSU)$ 
            \hfill \texttt{ // Recover Cross-Chain Transaction Intent}
        \STATE Update account states according to $DSU$
            \hfill \texttt{ // Update Account Balance State}
        \STATE $VisitedTxMap[ca] \gets VisitedTxMap[ca] \cup t$
            \hfill \texttt{ // Mark $t$ as visited for $ca$}
        \IF{$DSU$ satisfies expansion policy}
            \STATE $ST_a.push(DSU.r)$; 
                \hfill \texttt{ // Expand Illicit Money Flow Topology}
            \STATE \textbf{break}
        \ENDIF
    \ENDFOR
    \STATE Pop $ca$ if all its transactions are processed
\ENDWHILE
\STATE $(G, Des) \gets PostAnalysis()$
    \hfill \texttt{ // Perform Quantitative Post-Tracing Analysis}
\end{algorithmic}
\vspace{-0.1cm}
\end{algorithm}

After processing a transaction, AMLGuard evaluates whether the inferred DSU satisfies the forward expansion policy (L9). 
If the transaction is deemed to propagate illicit funds to a new address, the receiver address encoded in the DSU is pushed onto the stack, 
 and the algorithm immediately shifts focus to the newly discovered address by breaking the inner loop (L10). 
This depth-first expansion strategy prioritizes continuous tracing of fund flows along suspected laundering paths, enabling AMLGuard to follow the end-to-end movement of illicit assets. 
When no further expansion is triggered, AMLGuard continues analyzing the remaining transactions of the current address. Once all transactions associated with $ca$ have been processed, the address is removed from the stack (L14), and the algorithm resumes from the next address on the stack (L3). The iterative process ends when the address stack becomes empty. 
Upon termination, it outputs the reconstructed laundering topology, highlighting the final holders of illicit assets and their aggregation at exit services such as centralized exchanges. This semantic-guided workflow enables concise reconstruction of money laundering behaviors while avoiding redundant exploration and execution-level noise.

\section{Evaluation}

\noindent We aim to address the following research questions:

\begin{itemize}[leftmargin=*]
    \item \textbf{RQ1.} How effective is AMLGuard in tracking money laundering compared to the state-of-the-art methods?
    \item \textbf{RQ2.} What about the efficiency of AMLGuard?
    \item  \textbf{RQ3.} What are the effects and contributions of different components of AMLGuard in enhancing the accuracy and completeness of laundering tracking?
    \item \textbf{RQ4.} How helpful is AMLGuard in assisting manual audits for money laundering investigation?
\end{itemize}

\subsection{Experiment Setup}\label{experiment}


\noindent \textbf{Dataset.} Due to the lack of publicly available large-scale AML datasets for EVM blockchains, we constructed a comprehensive AML dataset to support research in crypto-native AML domain. 
Tracing the money laundering process of an attack is often labor-intensive and time-consuming. 
To accelerate this process, we collect detailed money laundering reports published before June 2025 by five reputable blockchain security firms: CertiK~\cite{Certik}, Beosin~\cite{Beosin}, SlowMist~\cite{SlowMist}, BlockSec~\cite{BlockSec}, and SharkTeam~\cite{SharkTeam}. 
These reports must include laundering tracing efforts and AML-related insights, providing valuable information for our dataset construction. 
As a result, we collected 82 security incidents, encompassing seven blockchain networks, e.g., ETH~\cite{Ethereum}, BSC~\cite{BSC}, and Arbitrum~\cite{Arbitrum}.

We involve three authors of this paper with expertise in blockchain security, each having two years of experience in blockchain security research. 
For each incident, three authors independently reconstruct the money laundering paths using blockchain explorers~\cite{Etherscan}, guided by the reported clues. To reduce annotation overhead and focus on important laundering paths, transfer paths with negligible value are excluded. 
The independently reconstructed paths are then jointly consolidated through joint review into a unified laundering trace. 
During this process, we further leveraged the decompilation provided by blockchain explorers and cross-chain protocol documentation to semantically recover interrupted cross-chain transactions.
To evaluate the consistency of the reconstructed laundering traces, we compute Cohen's Kappa coefficient among the three authors based on their independent annotations. The obtained Kappa score is $K=0.9320$, indicating a strong agreement among annotators and demonstrating the reliability of our dataset construction process.
As a final verification step, we cross-check the reconstructed paths using \textbf{MetaSleuth}~\cite{MetaSleuth}, a commercial AML tool developed by BlockSec~\cite{BlockSec}. 
While MetaSleuth may miss certain flows and cannot automatically recover complete laundering processes, it provides a reliable reference for cross-checking. Following this procedure, we ultimately obtained 82 high-quality laundering traces, consisting of 63 single-chain datasets ($D_s$) and 19 EVM cross-chain datasets ($D_c$). 


\textbf{DeFi Operation Knowledge Base Construction.} To enable implicit DeFi semantic inference, AMLGuard builds a high-quality DeFi Operation Knowledge Base covering DeFi protocols most frequently leveraged in real-world crypto money laundering.
Guided by total value locked (TVL) from DeFiLlama~\cite{DefiLlama}, we focus on four dominant protocol categories, i.e., DEXs, Cross-chain, staking protocols, and lending protocols, and select five representative EVM-compatible protocols from each category. 
For each protocol, we first identify its core logic contracts and then randomly sample 100 on-chain business transactions per protocol for manual inspection.
Transactions sharing identical function signatures and argument patterns are deduplicated to reduce semantic redundancy. 
Each transaction is independently analyzed by two authors, extracting its essential semantic features as defined in Section~\ref{llm_based}.
Disagreements are resolved through joint discussion with a third author, ensuring annotation consistency and semantic accuracy.
This process yields 1,787 unique DeFi transactions with precise semantic labels, spanning common laundering-relevant operations across the four type protocols.

 \textbf{Implementation.}  
We implement AMLGuard using over 3K lines of Python code, built on top of the Scrapy crawler framework~\cite{alshammari2021data}.
Transaction and contract data are retrieved via blockchain explorer APIs~\cite{Etherscan} and Web3 API~\cite{Web3}.
For implicit DeFi semantic inference, AMLGuard adopts GPT-4o from OpenAI~\cite{OpenAI} as the default large language model. In the RAG pipeline, semantic similarities are computed using the \textit{text-embedding-3-small} model, with the top-2 most relevant candidates at each step. The maximum number of LLM-based inference refinement iterations is set to three.
All experiments were conducted on machines equipped with the Intel(R) Xeon(R) Platinum 8163 CPU and 1TB RAM running Ubuntu 18.04.6 LTS.

 \textbf{Evaluation Metrics.} We evaluate the effectiveness of money laundering methods from two complementary perspectives. First, \textbf{Coverage} evaluates the ability to capture actual laundering activities and potential under-reporting, quantified by metrics:
(1) Recall: the proportion of ground-truth laundering addresses that are successfully traced. (2) Destination Recall: the proportion of ground-truth destination addresses recovered at the end of the laundering trace. Second, \textbf{Filtering} evaluates the effectiveness of excluding irrelevant paths in the AML process.
We adopt (1) Destination Precision, defined as the proportion of identified destination addresses that are genuine laundering destinations, to assess the quality of the final tracing results. 
Directly measuring false positives is impractical, as our ground truth intentionally excludes minor paths and some baseline methods often produce large, noisy graphs with many intermediate nodes. In practice, AML results are presented as tracing graphs that require manual analyst verification, making compact illicit topologies highly desirable.
We therefore indirectly quantify potential false positives and over-expansion by metrics: (2) Average AML Nodes per Incident: the average number of addresses in tracing graph per incident. (3) Average Transactions per Incident: the average number of transactions analyzed per incident.


\subsection{RQ1: Effectiveness of AMLGuard}

To evaluate the effectiveness of AMLGuard, we compare its performance with five state-of-the-art transaction tracing methods that satisfy the following two conditions: 1) open-source availability or ease of replication; 2) suitability for automated large-scale experiments. 
Consequently, we included TRacer~\cite{wu2023tracer}, Haircut~\cite{moser2014towards}, Poison~\cite{moser2014towards}, APPR~\cite{andersen2006local}, and XBlockFlow~\cite{wu2023toward} for evaluation. The first four are open-source and tested directly on our datasets: TRacer employs ranking-based relevance estimation and supports DeFi swap tracing; Haircut and Poison are taint-based heuristics with proportional and biased propagation respectively; and APPR applies approximate personalized PageRank to bias graph search. Although XBlockFlow is closed-source, we faithfully replicate its approach based on their paper for evaluation. 
We exclude DenseFlow~\cite{lin2024denseflow} because it targets a fundamentally different setting: it assumes a pre-constructed laundering graph and focuses on extracting salient subgraphs, rather than tracing illicit flows from a given suspicious address.
Thus, we select the above five methods as baselines.

\begin{table*}[htbp]
\centering
\caption{Tracking Performance Comparison on Datasets $D_s$ and $D_c$ among Baselines}
\label{tab:results}
\scalebox{0.78}{
\begin{tabular}{lcccccccccc}
\toprule
\multirow{2}{*}{\textbf{Method}} & 
\multicolumn{5}{c}{$D_s$} & 
\multicolumn{5}{c}{$D_c$} \\
\cmidrule(lr){2-6} \cmidrule(lr){7-11}
 & Re & D-Re & D-Prec & Avg. Tx & Avg. Node
 & Re & D-Re & D-Prec & Avg. Tx & Avg. Node \\
\midrule
TRacer            & 74.5\% & 47.6\% & - &  167K    &  59K   & 9.8\% & 22.7\% & - &  202K   & 65K \\
Haircut           & 72.3\% & 43.9\% & - &  74K   &  27K   & 9.8\% & 18.2\% & - & 29K   & 14K \\
Poison            & 72.3\% & 37.8\% & - &  76K  &  29.6K   & 14.1\% & 22.7\% & - & 50K   & 21.6K  \\
APPR              & 86.5\% & 56.1\% & - &  228K   &  98K   & 14.1\% & 22.7\% & - &  215K  & 76K \\ 
XBlockFlow & 93.4\% & 87.7\% & 88.1\% & 214 & 38 & 48.5\% & 60.1\% &65.6\%  & 84 & 26 \\
\midrule
AMLGuard w/o SA   & 85.2\% & 80.7\% & 87.6\% &  221  &  28 & 42.3\% & 56.3\% & 63.4\% & 96   & 23 \\ 
AMLGuard w/o LBSA &  94.8\%  & 89.1\%  & 91.7\%  &  290 & 55 & 82.5\%  & 87.8\%  & 81.2\% & 174 & 36 \\
AMLGuard     & 98.4\% & 94.1\% & 94.4\%  & 299 & 44  & 95.8\% & 93.8\% & 87.6\% & 287 & 74 \\
\bottomrule
\end{tabular}}
\end{table*}

Table~\ref{tab:results} summarizes the overall tracing performance. Filtering metrics directly reflect the manual verification cost faced by security analysts. AMLGuard traces on average 299 transactions and 44 labeled nodes per incident on $D_s$ and 287 transactions and 74 nodes on $D_c$, while achieving a destination precision of 94.4\% and 87.6\%, respectively. 
Compared to other methods, AMLGuard maintains a compact illicit topology with substantially fewer false positives.
Although directly measuring precision for the four heuristic-based baselines is impractical, these methods trace tens of thousands of transactions and nodes per incident. This excessive expansion arises from their inability to correctly interpret DeFi semantics, causing frequent misclassification of benign DeFi contracts as laundering participants and introducing substantial irrelevant noise. 
Despite keeping the traced topology small, AMLGuard achieves the highest recall (98.4\%) and destination recall (94.1\%) on $D_s$, demonstrating its superior ability to identify laundering participants and trace the final destinations of illicit funds.
On the cross-chain laundering dataset $D_c$, AMLGuard maintains 95.8\% recall and 93.8\% destination recall, substantially outperforming the best baseline (48.5\% and 60.1\%, respectively). This result highlights the critical role of cross-chain semantic recovery in enabling continuous tracing across blockchains. 
Although XBlockFlow achieves relatively high recall and precision with fewer traced entities on $D_s$, its lack of cross-chain semantic recovery leads to inaccurate tracing results on $D_c$.

To better understand the root causes of false negatives, we manually analyze the incidents where incomplete laundering paths are observed, and summarize the main reasons as follows:  
1) In some cases, mule addresses grant approve permissions to centralized exchange (CEX) addresses, which then actively transferred tokens out. However, since AMLGuard currently treats CEX addresses as terminal exit services, it does not trace their subsequent internal operations. This could be improved in the future by incorporating passive token movement analysis. 2) To mitigate path explosion, when AMLGuard encounters addresses that acted as complex intermediaries, involved in multiple overlapping laundering flows, it excludes addresses whose outgoing transfer amount exceeds their incoming transfer amount based on a heuristic algorithm.
And upon further inspection, most false positives are also attributed to addresses involved in multiple laundering processed, which complicated the evaluation of their asset states.
Overall, these results highlights the ability of AMLGuard to interpret DeFi transaction semantic and recover cross-chain semantic, which enables precise and reliable anti-money laundering tracing.

\subsection{RQ2: Efficiency of AMLGuard}
\begin{figure}[htbp]
\centerline{\includegraphics[width=0.9\linewidth]{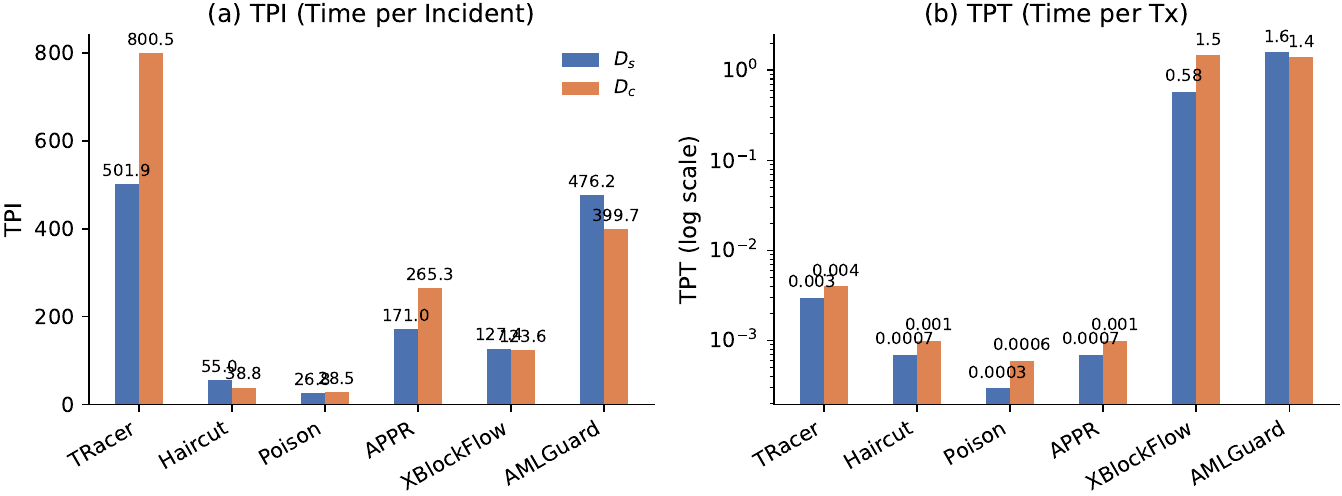}}
\caption{Efficiency Comparison on Datasets $D_s$ and $D_c$}
\label{tab: efficiency}
\vspace{-0.1cm}
\end{figure}

We evaluate the efficiency of AMLGuard by measuring its runtime in laundering path tracing. We report the \textbf{average analysis time per incident (TPI)} as an end-to-end efficiency metric.
As shown in Fig.~\ref{tab: efficiency}, AMLGuard achieves an average analysis time of 476.2 s per incident on $D_s$ and 399.7 s on $D_c$. In comparison, TRacer incurs significantly higher analysis time (501.9 s and 800.5 s), as it lacks accurate DeFi semantic understanding and consequently processes a large number of redundant interactions. 
Haircut, Poison, and APPR do not conduct semantic analysis. Instead, they rely on the graph-based heuristics algorithm, which leads to faster runtime but at the cost of lower accuracy and higher noise. XBlockFlow applies taint analysis and ignores complex DeFi transactions, which reduces the number of analyzed transactions (Table~\ref{tab:results}) and leads to lower analysis time. 

Compared to baseline methods, AMLGuard performs fine-grained transaction semantic analysis, which introduces additional time overhead. To quantify this overhead, we further evaluate the \textbf{average analysis time per transaction (TPT)}. As shown in Fig.~\ref{tab: efficiency}, AMLGuard spends 1.6 s per transaction on $D_s$ and 1.4 s on $D_c$, which is higher than the baseline methods.
This overhead stems from fine-grained DeFi semantic analysis. However, the additional cost is amortized by significantly reducing the total number of transaction analyzed, as shown in Table~\ref{tab:results}.
Overall, AMLGuard strikes a practical balance between tracing capability and computational efficiency, achieving accurate laundering path reconstruction within an acceptable processing time.

\subsection{RQ3: Component Contribution and Effect Analysis} \label{ablation}

\subsubsection{Ablation Study}
AMLGuard integrates three core components to enable transaction-level semantic analysis. 
To quantify the contribution of each component, we conduct an ablation study with two system variants. 
\textbf{Variant1 (AMLGuard w/o SA)} disables the entire semantic analysis pipeline for complex DeFi transactions. Only exit-related operations (e.g., CEX deposits) are explicitly processed, while the remaining tracing logic is unchanged. 
\textbf{Variant2 (AMLGuard w/o LBSA)} disables only the LLM-based semantic inference module. Transactions that cannot be interpreted by rule-based analysis are skipped without constructing semantic abstractions. 


The results are summarized in Table~\ref{tab:results}. On the dataset $D_s$, variant1 achieves 85.2\% recall and 80.7\% destination recall, while on the dataset $D_c$, the recall and destination recall drop sharply to 42.3\% and 56.3\%, respectively. 
Compared with the full system, this indicates that transaction-level semantic analysis is critical for uncovering hidden token transfer behaviors, particularly in cross-chain laundering scenarios. In addition, destination precision drops from 94.4\% to 87.6\% on $D_s$, and from 87.6\% to 63.4\% on $D_c$, indicating that insufficient semantic understanding leads to inaccurate destination identification.
Variant2 shows a milder but consistent degradation. 
Recall, destination recall, and destination precision drop to 94.8\%, 89.1\%, and 91.7\% on $D_s$, and to 82.5\%, 87.8\%, and 81.2\% on $D_c$. 
This suggests that while rule-based DeFi operation identification can capture common DeFi patterns, LLM-based semantic inference is critical for correctly interpreting complex and evolving DeFi transactions beyond the coverage of static rules.




\subsubsection{Effectiveness of LLM-based DeFi Operation Inference}
In our constructed AML dataset, approximately 77.1\% of transactions are resolved by static rules, while the remaining 22.9\% require LLM-based inference to identify implicit DeFi behaviors. Among the transactions requiring LLM-based inference, we find that 69.7\% of cases are correctly inferred in the first attempt. The remaining cases are resolved through iterative refinement, with 21.6\%, 7.3\%, and 1.4\% successfully addressed after one, two, and three additional iterations, respectively. These results demonstrate the effectiveness of LLM-based inference in resolving implicit DeFi behaviors within our dataset. 

However, since the involved protocols may overlap with the knowledge acquired during the inference process, we further evaluate whether LLM-based DeFi operation inference can generalize to previously unseen protocols. We conduct a controlled experiment on previously unseen protocols.
For each DeFi operation category, we select three representative protocols excluded from the RAG knowledge base and additionally include 1inch, a large-scale DEX aggregator supporting diverse and composite DeFi interactions. 
For each selected protocol, we collect all related transactions and randomly sample a total of 200 DeFi transactions, distributed proportionally across protocol types. The ground-truth DeFi operation labels and details are independently annotated by three authors following the same criteria as before.


\begin{table}[htbp]
\centering

\caption{Performance of LLM-based DeFi Operation Inference (Metrics: accuracy/time/monetary cost)}
\makebox[\linewidth][c]{
\scalebox{0.82}{
\begin{tabular}{ccccc}
\bottomrule
\textbf{Metrics} & \textit{Pure} & \textit{w/o Feedback} & \textit{w/o RAG} & \textit{Full} \\
\hline
gpt-4o & 67.0\% / 3.2s / 0.0056\$ & 83.5\% / 4.9s / 0.0059\$  &  86.5\% / 4.8s / 0.0080\$  & 97.0\% / 5.8s / 0.0074\$\\

deepseek-v3 &  71.5\% / 3.6s / 0.0054\$ & 87.0\% / 4.8s / 0.0055\$ & 83.5\% / 4.9s / 0.0074\$  & 96.5\% / 5.6s / 0.0064\$ \\

gemini-2.5-flash & 67.0\% / 4.3s / 0.0053\$ & 88.5\% / 5.8s / 0.0054\$ & 85.5\% / 5.7s / 0.0085\$ \ & 97.5\% / 6.5s / 0.0064\$  \\
\bottomrule
\end{tabular}} }
\label{tab:operation inference}
\end{table}

We evaluate three representative LLMs: \textit{gpt-4o}, \textit{deepseek-v3}, and \textit{gemini-2.5-flash}, under identical settings. For each model, we further compare four inference variants: \textit{Pure}, which relies solely on the LLM for operation inference; \textit{w/o Feedback}, which removes the feedback mechanism; \textit{w/o RAG}, which disables retrieval-augmented context; and \textit{Full}, which adopts the complete AMLGuard design. 
In addition to inference accuracy, we also measure the corresponding inference time overhead and monetary cost of different configurations. The results are shown in Table~\ref{tab:operation inference}.

Under the \textit{Full} configuration, all models achieve over 96\% accuracy , with average inference time of 5.8s, 5.6s and 6.5s, and an average monetary cost of approximately \$ 0.007 per tx. 
In practice, both the time and monetary overhead are negligible, as laundering campaigns typical interact with the same protocols repeatedly. 
Once an operation is inferred, AMLGuard constructs and caches the corresponding operation signature, eliminating redundant LLM invocations for subsequent transactions.
Across all three LLMs, the ablation results exhibit consistent trends. Taking \textit{gpt-4o} as a representative example, the \textit{Pure} variant, which relies solely on the LLM, achieves only 67\% accuracy, indicating that raw transaction data alone is insufficient for reliable semantic inference. Removing the feedback mechanism (\textit{w/o Feedback}) leads to a substantial drop from the full configuration (97.0\% to 83.5\%), highlighting the importance of iterative semantic validation in correcting ambiguous or partially inferred operations. Similarly, disabling retrieval-augmented context (\textit{w/o RAG}) further degrades accuracy to 86.5\%, demonstrating that in-context DeFi examples play a critical role in guiding the LLM toward correct intent interpretation.

\subsubsection{Effectiveness of Cross-chain Semantic Recovery}

\begin{table}[htbp]
\centering
\caption{Performance of Cross-Chain Semantic Recovery}
\makebox[\linewidth][c]{
\scalebox{0.78}{
\begin{tabular}{ccccc @{\hspace{1.2em}} ccccc}
\hline
\textbf{Incident} & \textbf{From} & \textbf{To} & \textbf{Tx} & \textbf{Succ} &
\textbf{Incident} & \textbf{From} & \textbf{To} & \textbf{Tx} & \textbf{Succ} \\
\hline
Wault.Finance & BSC & ETH & 3 & 3 &
XKingdom & Arbitrum & ETH & 5 & 5 \\

QBridge & BSC & ETH & 51 & 51 &
Li.Fi & ETH & Arbitrum & 6 & 3 \\

Paraluni & BSC & ETH & 5 & 5 &
SenecaUSD & ETH & Arbitrum & 15 & 15 \\

New Free DAO & BSC & ETH & 2 & 2 &
WooPPV2 & Arbitrum & ETH & 29 & 28 \\

BSC Token Hub & BSC & ETH & 46 & 46 &
Sonne Finance & Optimism & ETH, Arbitrum & 52 & 47 \\

CirculateBUSD & BSC & ETH & 3 & 3 &
UtopiaSphere & BSC & ETH & 10 & 4 \\

Chibi Finance & Arbitrum & ETH & 2 & 2 &
DeltaPrime & Arbitrum & ETH & 2 & 2 \\

Exactly Protocol & Optimism & ETH & 5 & 5 &
Clober DEX & BASE & ETH & 13 & 7 \\

OKK DEX & ETH & BSC, Polygon & 93 & 91 &
Magic & Arbitum & ETH & 13 & 13 \\

Radiant Capital & Arbitrum & ETH & 93 & 93 &
\textbf{Total} & - & - & \textbf{448} & \textbf{424} \\

\hline
\end{tabular}} }
\label{tab:crosschain-performance}
\end{table}

To evaluate the cross-chain semantic recovery capability, we further analyze the cross-chain transactions involved in each incident of the dataset $D_c$. The results are summarized in Table~\ref{tab:crosschain-performance}. Overall, AMLGuard successfully recovers 424 out of 448 cross-chain transactions, achieving a recovery accuracy of 94.6\%. The remaining failures are mainly caused by proxy-based cross-chain contracts, where the logical contract invoked during analysis differs the logic contract used at the time of laundering. In future work, we plan to incorporate historical slot-state analysis to accurately locate the logic contract and improve recovery accuracy.

\subsection{RQ4: Case Study}
To evaluate how AMLGuard assists auditors in real-world AML investigations, we conduct a comparative user study under three settings: Expert Only, Expert with MetaSleuth~\cite{MetaSleuth} and Expert with AMLGuard. In both settings, experts are allowed to use blockchain explorers to assist their investigation. 
We select two recent real-world incidents as case studies: the Li.Fi incident~\cite{Li.Fi} (July 2024, $\$$11.6 M loss) and the 0xInfini incident~\cite{0xInfini} (February 2025, $\$$49 M loss). 
We recruit six auditors with at least two years of auditing experience and extensive expertise in analyzing transaction behaviors. Each incidents is investigated in two rounds under the three experimental settings.

\begin{figure}[htbp]
\centerline{\includegraphics[width=0.55\linewidth]{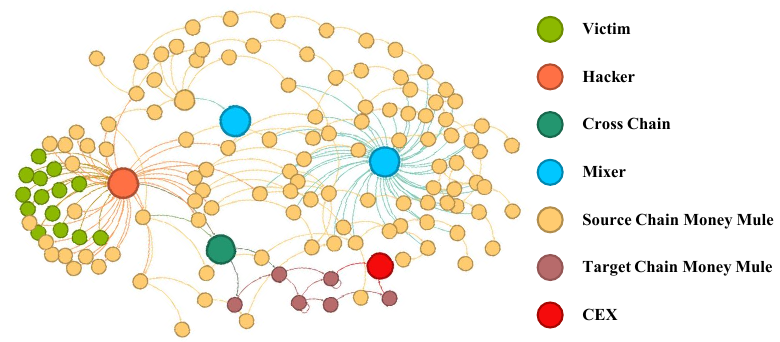}}
\caption{Tracing Visualization for Li.Fi Incident}
\label{case study}
\vspace{-0.2cm}
\end{figure}

Under the Expert Only setting, auditors rely solely on blockchain explorers and spend approximately 25 mins (Li.Fi) and 8 mins (0xInfini) manually reconstructing the laundering process. With MetaSleuth assistance, investigation time is reduced to 21 mins and 5 mins, respectively; however, the lack of precise account-state tracking and the inability to support cross-chain tracing still impose significant manual effort.
In contrast, under the Expert with AMLGuard setting, auditors complete the investigations in just 10 mins and 3 mins, including AMLGuard’s execution time, corresponding to a manual effort reduction of over 60\%.
Taking the Li.Fi incident as an example, the hacker employed over 100 intermediary addresses for multi-hop fund transfers, involving numerous token swaps and cross-chain transfers. Existing tracing tools fail in this setting, as they neither interpret complex DeFi semantics nor recover cross-chain intent, and thus generate large, noisy transaction graphs that are impractical for manual analysis.
The auditor had to manually analyze complex DeFi operations and reconstruct the path step-by-step. 
In contrast, AMLGuard automatically inferred high-level DeFi semantics, recovers cross-chain transfer intent, and presented an end-to-end flow of the stolen assets. The reconstructed laundering graph by AMLGuard is shown in Fig.~\ref{case study}. 
The results demonstrate that AMLGuard can effectively support real-world AML investigations by reducing analysis time and effort, thereby enabling regulators to take timely actions.

\section{Discussion}\label{discussion}


\noindent \textbf{Internal Validity.} Our ground-truth dataset is constructed based on manual analysis, which may introduce potential inaccuracies. The accuracy of the annotated laundering paths directly impacts the effectiveness of AMLGuard. 
To mitigate this, we perform double verification of each case and cross-validate the results using MetaSleuth~\cite{MetaSleuth}. Furthermore, AMLGuard operates on account-based blockchains, where token balances are aggregated at the account level. When illicit funds enter an account that holds benign assets, subsequent outgoing transfers cannot be unambiguously attributed to the laundered funds. This limitation stems from the inherent ambiguity of the account-based execution model rather than the implementation of AMLGuard.

\noindent \textbf{External Validity.} 
Our framework focuses on laundering behavior observable on EVM-compatible chains up to known exit services, such as CEXs and coin mixers. 
In our ground-truth dataset, 66 out of 82 laundering incidents involve partial fund flows into mixers, demonstrating that mixer-related laundering represents a significant portion of real-world cases. 
However, real-world AML investigations may need to track laundering activities beyond these observable exit services, particularly after funds are routed through mixers. 
While these behaviors fall outside the scope of this paper, prior research has explored such threats in depth~\cite{buterin2024blockchain, wang2023zero,gilbert2024unlocking,de2024guideenricher,du2023breaking}. 
Importantly, our framework is modular by design and can be extended to incorporate specialized tracking for mixer-based laundering.

\noindent \textbf{Scope and Future Directions.}
While AMLGuard focuses on forward tracing from known illicit sources to identify laundering flows, backward propagation from suspicious addresses or exit services can provide complementary evidence for AML investigations. 
Through further analysis, we identify backward propagation as a promising yet challenging direction for improving AML investigation. 
Specifically, two key challenges remain: (1) how to accurately associate destination-chain transactions with their corresponding source-chain activities, especially when internal transactions lack sufficient information; and (2) how to define appropriate stopping criteria for backward tracing, particularly when the tracing process reaches benign entities or victim protocols. 
Addressing these challenges and developing comprehensive backward propagation techniques represent important directions for future work.

In addition, some DeFi operations may span multiple transactions rather than being completed within a single atomic transaction. 
Such multi-transaction workflows introduce additional challenges for operation inference because their semantics depend on the interactions among multiple intermediate states. 
However, compared with single-transaction operations, multi-transaction workflows provide weaker atomicity guarantees and expose intermediate states publicly, making them potentially more vulnerable to front-running and MEV attacks. 
Moreover, in our analysis of real-world laundering cases, we do not observe such complex DeFi operations being exploited in practice. 
Therefore, AMLGuard currently focuses on transaction-level DeFi operation inference, while extending the framework to support more complex multi-transaction DeFi workflows remains an interesting direction.

\section{Related Work}
 \textbf{Crypto-native Anti-Money Laundering.}
In blockchain systems, pseudonymous accounts and real-world identities are usually unlinked. In the cryptocurrency world, the first publicly available dataset related to ML is the Elliptic dataset~\cite{weber2019anti}, which consists of a transaction network formed by Bitcoin transactions, categorized into licit and illicit transactions. Several studies have applied different AML techniques to identify illicit fund flows~\cite{weber2019anti, lorenz2020machine,wu2021towards, altman2023realistic, alarab2020competence,vassallo2021application}. Furthermore, many studies have also focused on addressing AML challenges in account-based blockchain platforms~\cite{lin2025connector,lin2025track,buterin2024blockchain,song2024illicit, farrugia2020detection, liu2023graph}. Wu et al.~\cite{wu2023tracer} introduce an innovative personalized ranking method for effective transaction tracing on account-based blockchains. And Wu et al.~\cite{wu2023toward} construct the first crypto-asset ML dataset on the Ethereum blockchain platform and conduct a comprehensive analysis for main stages of blockchain ML. Lin et al.~\cite{lin2024denseflow} design the suspiciousness metric for accounts and transactions based on the traits of ML behavior and trace ML activities by finding dense subgraphs. 

\noindent \textbf{Transaction Behavior Analysis.}
Smart contracts execute their functions by interacting with diverse transactions that include different actions and semantics~\cite{kong2025smart,liu2024using,wu2025detecting}. Recently, many studies have sought to analyze the transaction behaviors of smart contracts to enhance blockchain security~\cite{wu2023know, xie2024defort, zhong2025defiscope, wu2024tokenscout,wu2025safeguarding,aquilina2021etherclue,pan2024ethershield}. Zhang et al.~\cite{zhang2020txspector} replay history transactions and record EVM bytecode-level traces, utilizing some pre-defined rules to detect logic vulnerabilities. DeFiRanger~\cite{DeFiRanger} constructs cash flow trees from transaction sequences, lifts the semantics of trees to high-level DeFi actions and employs specific patterns to identify price manipulation. SPCon~\cite{liu2022finding} mines past history transactions of a contract to recover a likely access control model, which can be checked against information policies, identifying access control bugs. DeFiWarder~\cite{su2023defiwarder} also replays history transactions and performs role mining to infer the semantics of different accounts, detecting abnormal token leakage vulnerabilities. Zhang et al.~\cite{zhang2023your} analyze attacking transactions to extract structured attack patterns and synthesize counterattack smart contracts capable of front-run the attacks.


\section{Conclusion}

We propose AMLGuard, a semantic-aware anti-money laundering framework that enables tracing of illicit funds on account-based blockchains. Given a complex DeFi transaction, AMLGuard performs transaction semantic analysis to infer its underlying semantics, tracing complex money laundering activities.
Evaluation on 82 real-world laundering cases shows that AMLGuard can produce compact illicit fund-flow topologies, while achieving high address recall and precision.

\section{Data Availability}
Our replication package is available online: \url{https://figshare.com/s/e01b691346bb352701e5}.

\begin{acks}
This work was supported by National Natural Science Foundation of China (62372367, 62232014, 62272377, 62372368), and Shaanxi Province Sanqin Talent Introduction Program.
\end{acks}

\bibliographystyle{ACM-Reference-Format}
\bibliography{AMLGuard}










\end{document}